\documentclass{elsart3p}

\usepackage{natbib}
 \usepackage{graphicx}

\usepackage{amssymb}
\usepackage{ulem}

\usepackage{lineno}

\usepackage{amsmath}

\newcommand{\ms}{\mbox {$\rm M_{\odot}$}}
\newcommand{\ace}{\mbox {$\alpha_{\rm CE}$}}
\newcommand{\md}{\mbox {$\dot{M}$}}

\newcommand{\myr}{\mbox {~${\rm M_{\odot}~yr^{-1}}$}}
\newcommand{\es}{\mbox {~erg s$^{-1}$}}
\newcommand{\al}{\mbox {$\alpha_{\rm CE} \times \lambda$}}
\newcommand{\te}{\mbox {$T_{\rm eff}$}}
\newcommand{\per}{\mbox {$P_{\rm orb}$}}

\def\aap{A\&A }
\def\aa{A\&A}
\def\araa{Ann. Rev. Astron. and Astrophys. }
\def\apj{ApJ }
\def\apjl{ApJ }
\def\mnras{MNRAS }

\def\apgt{\ {\raise-.5ex\hbox{$\buildrel>\over\sim$}}\ }
\def\aplt{\ {\raise-.5ex\hbox{$\buildrel<\over\sim$}}\ }

\def\spose#1{\hbox to 0pt{#1\hss}}
\def\simless{\mathrel{\spose{\lower 3pt\hbox{$\mathchar"218$}}
        \raise 2.0pt\hbox{$\mathchar"13C$}}}
\def\simgreat{\mathrel{\spose{\lower 3pt\hbox{$\mathchar"218$}}
        \raise 2.0pt\hbox{$\mathchar"13E$}}}
\def\lta{\mathrel{\spose{\lower 3pt\hbox{$\mathchar"218$}}
        \raise 2.0pt\hbox{$\mathchar"13C$}}}
\def\gta{\mathrel{\spose{\lower 3pt\hbox{$\mathchar"218$}}
        \raise 2.0pt\hbox{$\mathchar"13E$}}}

\begin{document}

\begin{frontmatter}



\title{Evolution of low-mass binaries with
black-hole components}


\author[Moscow,Paris]{L. Yungelson\corauthref{cor}},
\corauth[cor]{Corresponding author.}
\ead{lry@inasan.ru}
\author[Paris,Polska]{J.-P. Lasota}
\address[Moscow]{Institute of Astronomy of the Russian Academy of
            Sciences, 48 Pyatniskaya Str., 119017 Moscow, Russia}
\address[Paris]{Institut d'Astrophysique de Paris, UMR 7095 CNRS,
           Universit\'e Pierre et Marie Curie, 98bis Bd Arago, 75014 Paris,
           France}
\address[Polska]{Astronomical Observatory, Jagiellonian University, ul. Orla 171, 30-244 Krak\'ow, Poland}
\ead{lasota@iap.fr}

\begin{abstract}
We consider evolutionary models for the population of short-period
($\per \lta 10$\,hr) low-mass black-hole binaries (LMBHB) and
compare them with observations of soft X-ray transients (SXT).
Evolution of LMBHB is determined by nuclear evolution of the donors
and/or orbital angular momentum loss due to magnetic braking by the
stellar wind of the donors and gravitational wave radiation. We show
that the absence of observed stable luminous LMBHB implies that
upon RLOF by the low-mass donor angular momentum losses are
substantially reduced with respect to the Verbunt \& Zwaan
``standard'' prescription for magnetic braking. Under this
assumption masses and effective temperatures of the model
secondaries of LMBHB are in a satisfactory agreement with the masses
and effective temperatures (as inferred from their spectra) of the
observed donors in LMBHB. Theoretical mass-transfer rates in SXTs
are consistent with the observed ones if one assumes that accretion
discs in these systems are truncated (``leaky''). We find that the
population of short-period SXT is formed mainly by systems which had
unevolved or slightly evolved ($X_c \gta 0.35$) donors at the
Roche-lobe overflow. Longer period ($\per \simeq (0.5 - 1)$\
day) SXT might descend from systems with initial donor mass about
1\,\ms\ and $X_c \lta 0.35$. It is unnecessary to invoke donors with
almost hydrogen-depleted cores to explain the origin of LMBHB. Our
models suggest that a very high efficiency of common-envelopes
ejection is necessary to form LMBHB, unless currently commonly
accepted empirical estimates of mass-loss rates by winds for pre-WR
and WR-stars are significantly over-evaluated.
\end{abstract}


\end{frontmatter}

\section{Introduction}
\label{sec:intro} Out of twenty dynamically-confirmed black-hole
candidate X-ray binaries ten have K/M spectral type secondaries and
orbital periods $\aplt$1 day \citep{2006ARA&A..44...49R}. All these
systems are transient in X-ray. Their quiescent X-ray luminosity is
$\simeq (3\,10^{30} -- 3\,10^{33})$\,\es, while in the outburst
luminosity may be by a factor $\sim 10^5 - 10^8$ higher. These low-mass black-hole binary
(LMBHB) X-ray systems belong to the class of ``Soft X-ray
transients'' (SXT). Recent reviews of observational data on SXT may
be found e.g. in \citet{2006csxs.book..215C,2006ARA&A..44...49R}.
From the evolutionary stand-point SXT are considered as semidetached
binaries in which matter is transferred to the compact object via
accretion disc. The commonly accepted model for the variability
of SXT is based on the thermal-viscous instability of irradiated
accretion discs \citep[see ][and references therein]{lasota01}. The
estimated number of SXT in the Galaxy ranges from several hundred
\citep{csl97} to several thousand \citep{romani1998}.
\begin{figure}[ht!]
\includegraphics[angle=-90,width=0.48\textwidth]{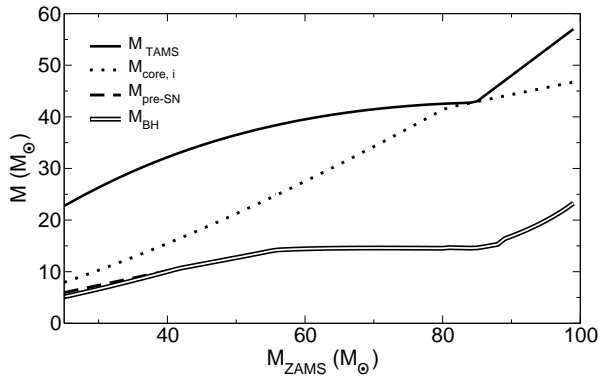}
\caption[]{Initial -- final mass relations for massive stars in
binaries. The mass of stars at terminal
main-sequence (TAMS), the initial mass of helium core ${\rm M}_{\rm
core, i}$, the pre-supernova mass ${\rm M}_{\rm pre-SN}$, and the
black-hole mass ${\rm M}_{\rm BH}$ are shown.}
\label{fig:imfm}
\end{figure}
Below, we will describe the models of a population of short-period
LMBHB, discuss their observational properties, and consider
possible existence of a population of faint LMBHB with quasi-stable,
cold accretion discs \citep{mnl99}. This review is based mainly on
the studies by \citet{yungelson_bh06} (henceforth, Paper I) and
Yungelson \& Lasota (in prep., Paper II).

\section{The model}
\label{sec:model}

\subsection{Formation of black holes with low-mass companions}
\label{sec:bh+ms}

Modeling of the Galactic population of LMBHB includes two major
steps: (a) modeling of the population of zero-age binaries
containing a black hole accompanied by a low-mass companion and (b)
following the evolution of the system to Hubble time.

The population of zero-age LMBHB was computed (Paper I)  using population synthesis code \textsf{SEBA} \\
\citep{pv96,py98,nyp+01,nyp04} with 250000 initial binaries with
$M_{10} \geq 25$\,\ms. Assumed time- and position-dependent Galactic
star formation history is based on the model of \citet{bp99}, but
for the inner 3 kpc of the Galaxy we doubled the star formation rate
given in the latter study to mimic Galactic bulge. We assumed a
50\% binarity rate (2/3 of stars in binaries), an IMF after
\citet{ktg03}, an initial distribution of separations in binaries
($a$) flat in $\log a$ between contact and 10$^6$ R$_\odot$, a flat
mass ratio distribution, and initial distribution of eccentricities
of orbits $\Xi(e)=2e$.

A summary of the assumptions on transformations of stellar masses in
the course of the evolution is shown in Fig.~\ref{fig:imfm}. Black
hole progenitors have $M_0=25 - 100$\,\ms.
The algorithm for formation of black holes in the code follows the
fall-back scenario \citep{fryer_kal_bh01} under assumption that
explosion energy is fixed at $10^{50}$ ergs, which is within the
expected range but favours formation of rather massive black holes
(up to $\sim15$\,\ms). Nascent black holes receive an asymmetric
kick at formation following \citet{pac90} velocity distribution with
$\sigma_v=300$ km~s$^{-1}$, but scaled down with the ratio of the
black hole mass to a neutron star mass. The population of LMBHB is
not sensitive to the assumed kick distribution, since scaled down
with $M_{\rm bh}/M_{\rm ns}$-ratio kick velocities are too small to
disrupt close binaries in SN explosions. Test run for Maxwellian
kick distribution for pulsars with $\sigma=265$\,km/s
\citep{hobbs2005} confirmed this assumption.
 For the
common-envelope phase we used the standard prescription
\citep{web84,khp87}, with the efficiency and structure parameters
$\ace$ and $\lambda$ product $\al= 2$ (see discussion of the issue
of \al\ in \S \ref{sec:concl}). For more detailed description of
input parameters we refer the reader to Paper I.

\subsection{Angular momentum loss}
\label{sec:aml}

Since low-mass components in LMBHB have KV/MV spectral types, it is
usually assumed that evolution of LMBHB is governed by angular
momentum losses (AML), like for cataclysmic variables. In our
computations we assumed that AML via magnetic stellar wind (MSW)
follows the model suggested by \citet{vz81}:
\begin{equation}
\dfrac{d\ln J}{dt} = -\dfrac{1}{2}\times 10^{-28}k^{2}\left(\dfrac{2\pi}{P}\right)^{10/3}\dfrac{1}{G^{2/3}}
\dfrac{M_t^{1/3}R_2^{4}}{M_1f^{2}},
\label{eq:msw}
\end{equation}
where $M_1$ is mass of the primary, $M_t$ -- total mass of the
system, $P$ -- orbital period, $R$ -- radius of the secondary, $k^2
\sim 0.1$ -- its gyration radius, $f \sim 1$ -- a parameter derived
from observations. Note, that this model extrapolates via almost two
orders in magnitude stellar rotation braking law derived by
\citet{sku72} for \textit{single} field stars to rapidly rotating
components of close binaries and assumes efficient spin-orbital
coupling.

Angular momentum loss via gravitational waves radiation (GWR) is described by the standard \citet{ll71} formula
\begin{equation}
\dfrac{d\ln J}{dt} = -\dfrac{32}{5}\dfrac{G^{5/3}}{c^5}\left(\dfrac{2\pi}{P}\right)^{8/3}
\dfrac{M_1\,M_2}{(M_1+M_2)^{1/3}}.
\label{eq:gwr}
\end{equation}

Below, we will refer to the model in which semidetached systems
obey Eqs.~(\ref{eq:msw}) and (\ref{eq:gwr}) as to a ``standard''
model.

\subsection{Population of unevolved LMBHB}
\label{sec:lmbhb}

\begin{figure}
\center
\includegraphics[angle=-90,width=\columnwidth]{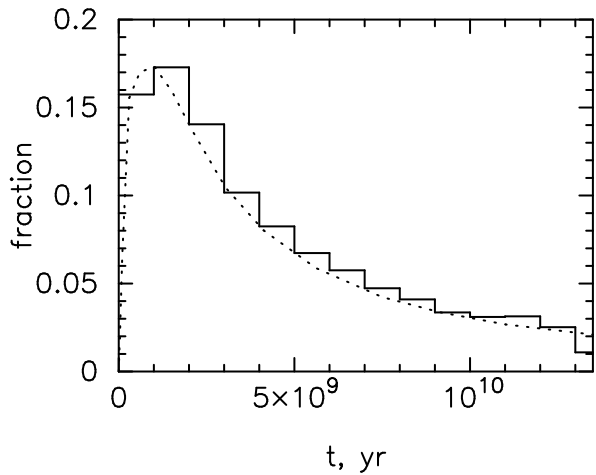}
\caption[]{Fractions of low-mass bh+ms binaries formed at different
epochs in the lifetime of the Galaxy (solid line) and the shape of
star formation rate (SFR) adopted in present study (dashed line). At maximum, the SFR is $\simeq 15.6$\,\myr. Current SFR is $\simeq 1.9$\,\myr.}
 \label{fig:history}
\end{figure}

\begin{figure}
\center
\includegraphics[width=\columnwidth]{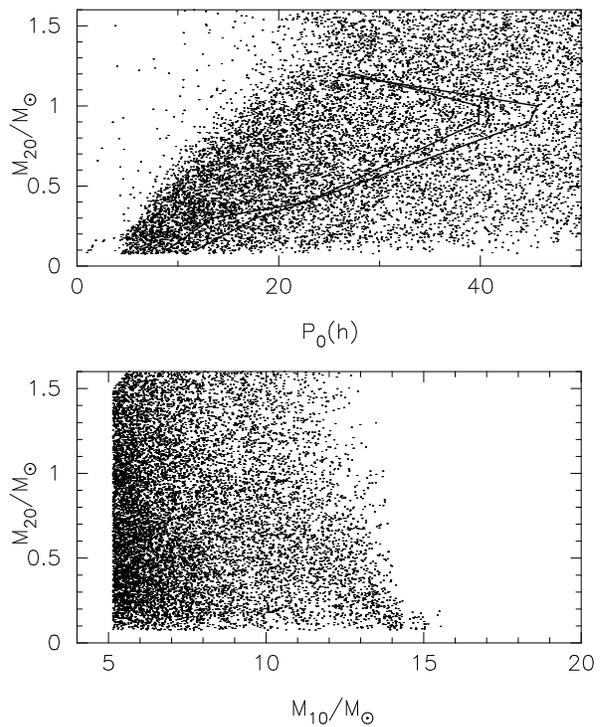}
\caption[]{Upper panel: relation between orbital periods and masses
of main-sequence companions of black holes after circularization of
the orbits. Systems with main-sequence components that may fill
their Roche lobes in their main-sequence lifetime or in the Galactic
disc lifetime (13.5 Gyr) are located to the left of the solid
curves that correspond to $M_{\rm
BH}=4\,\ms$ (left one) and $12\,\ms$ (right one), respectively. Lower panel: relation between initial
masses of components in bh+ms systems.}
\label{fig:p0m20}
\end{figure}

With our assumptions we find that, within a Hubble time (13.5 Gyr),
$\sim 49\,000$ binaries that have orbital periods below 2.0 day and
contain black holes accompanied by main-sequence stars with $M_{20}
<1.6$\,\ms\ (for which magnetic braking is supposed to operate)
were formed in the Galaxy.\footnote{This number is $\sim$0.01\% of
all binaries with black-hole components formed in Hubble time.}
Out of them, $\sim 17000$ were brought into contact by AML via MSW
and/or GWR and $\sim 14\,000$ evolved to shorter periods under
``standard'' assumptions on MSW (see \S~\ref{sec:aml}).

Because black holes form in the first $\sim$3 Myr after their
progenitor formation, formation history of bh+ms binaries
 strictly follows the star formation history (Fig.~\ref{fig:history}).
Figure \ref{fig:p0m20} shows relations between initial masses of
components in model systems and their post-circularization orbital
period\footnote{Every dot in the plot actually represents four or
five systems with similar combination of $M_{10}$, $M_{20}$, $P_0$,
but born at different time. Since we present one random realization
of the model, all numbers given are subject to Poisson noise.}.
Solid lines in the upper panel of the Figure outline the mass and
period range from which systems may be brought in the contact via
action of AML, however, in ``standard'' model only systems that
overflow Roche lobes at periods less than $\simeq 15$~hr evolve to
shorter periods (the rest have small helium cores and evolve to
longer periods, forming at the end helium white dwarfs, if time span
between contact and Hubble time is large enough; otherwise, at
present their donors are subgiants).

\section{Evolution of LMBHB}
\label{sec:evol}

For simulation of the current Galactic population of semidetached
LMBHB  we convolved the above-described ``underlying''
population of binaries born at different epochs in the history of
the Galaxy with the grids of evolutionary tracks for low-mass
components in the binaries with different combinations of masses of
components and post-circularization (initial) orbital periods. For
evolutionary computations, appropriately modified TWIN version
(September 2003) of  \citet{egg71} evolutionary code was
used. 
Every ``initial'' 
system was ``evolved'' 
through $13.5\,{\rm Gyr} - T_0$, where $T_0$ is the  
moment of formation.
In the Figures presented below, we show only systems that have at present
 $q=M_2/M_1 \geq 0.02$. At $q \aplt 0.02$ the
circularization radius of accretion stream becomes greater than the
outer radius of the accretion disc, resonance phenomena in the disc
become important and, though computations may be continued in the
same fashion as before, until models cease to converge due to very
low mass, actually it is unclear how mass transfer occurs (see Paper
I for more details). The systems with $q < 0.02$ have $P_{\rm orb}
\aplt 2$ hr and mass-transfer rates ($ < 10^{-10}$\,\myr). In the
``standard'' model currently more than 75\% of all low-mass ms+bh
systems have $q < 0.02$ and have yet to be observed if mass-transfer
occurs in them.

\subsection{Luminous persistent LMBHB?}
\label{sec:stable}

\begin{figure}[t!] \includegraphics[angle=-90,width=0.48\textwidth]{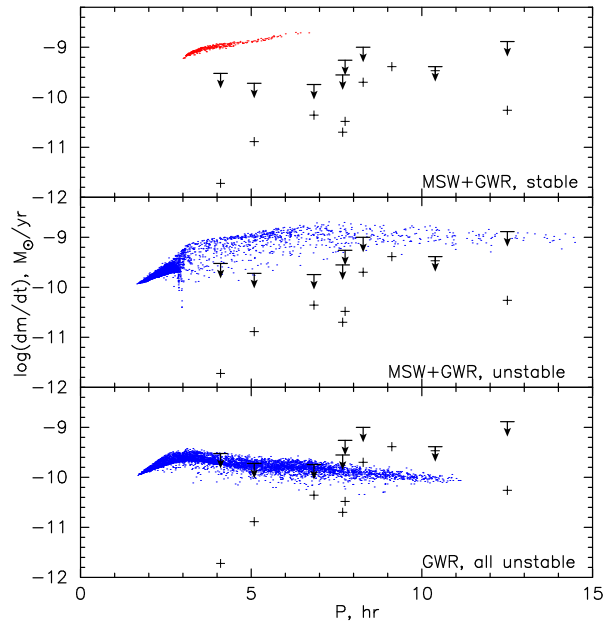}
\caption[]{Mass transfer rates vs. orbital period for the model population of
semidetached low-mass binaries with assumed black hole accretors. Upper panel
-- stable systems in the model for a ``standard'' Verbunt-Zwaan assumption on
AML via MSW. Middle panel -- population of unstable systems in the same
model. Lower panel -- population for the model in which MSW does not operate
in semidetached systems. In this model all binaries have unstable discs.
The arrows mark approximate upper limits
to the average mass-transfer rates for SXT under assumption of ``leaky'' discs,
while plusses show estimates of rates based on assumed recurrence period of 30
years, except for V616 Mon and IL~Lup whose recurrence times are known to be
$\sim 50$ and 10 years, respectively (see \S~\ref{sec:dmdt} for details).
}
\label{fig:dim}
\end{figure}

Figure \ref{fig:dim} shows the distribution of systems with $q >
0.02$ in the ``standard'' model over mass-transfer rates. The
upper two panels show the break-down of the binaries that have
stable or unstable accretion discs (370 and 2900 objects,
respectively). Irradiated discs are hot and stable if accretion rate
(in \myr) exceeds \citep{dubhamlas99}
\begin{equation}
\label{eq:mcrit} \md^+_{\rm crit} \approx 2.4 \times 10^{-11} M_{\rm
BH}^{-0.4} R_{d,10}^{2.1} \left(
\frac {\mathcal{C}}{5 \times 10^{-4}}\right)^{-0.5},
\end{equation}
where $M_{\rm BH}$ is the mass of the accretor in \ms,
$R_{d,10}$ is the disc outer radius in $10^{10}$ cm, and
$\mathcal{C}$ is a measure of heating of the disc by X-rays
\citep{shaksun73}; for a given disc radius $R$ the irradiation
temperature $T_{\rm irr}$ is
\begin{equation}
\label{eq:tirr} T_{\rm irr}^4=\mathcal{C} \frac{\md c^2} {4 \pi
\sigma R^2},
\end{equation}
where $\sigma$ is the Stefan-Boltzmann constant. We use
$\mathcal{C}=5 \times 10^{-4}$, following \citet{dubhamlas99} who
have found that this value is consistent with properties of
persistent low-mass X-ray sources, but actually
 $\mathcal{C}$ might e.g.
vary in time \citep{elh}.

However, no persistent LMBHB with $P_{\rm orb} \aplt 4$ hr are
observed\footnote{In fact no LMBHB at all are observed below $P_{\rm
orb} \sim 4$ hr.}. Stable model systems have an average
luminosity $L=0.1 \dot{M}c^2 \approx 5 \cdot 10^{36}$\es. The
accretion luminosity may be assumed to be emitted as that of a 1~keV
blackbody, corresponding to the ``high/soft'' X-ray spectral state
of LMBHB. Given their location in the Galaxy provided by the
population synthesis code, we may roughly estimate intervening X-ray
absorption, We found that the number of persistent systems expected
above the completeness limit for the X-ray sources seen by the
\textit{ RXTE All Sky Monitor} (about 3 mCrab or 0.2 counts/s) in
the 2-10~keV band is 85. Setting the threshold ten times higher
still yields 13 bright, persistent LMBHB with \textit{ASM} fluxes
comparable to e.g. SMC X-1. Noticeable number of stable ``observed''
persistent model sources remain in the model also under another
assumptions on the spectrum of their emission (see for details Paper
I). If the discs were not irradiated, no systems would be
stable. However, both persistent neutron-star low-mass X-ray
binaries and outbursting black hole transients show clear
signatures of irradiation. Note that with parameter
$\mathcal{C}=5\times 10^{-4}$, only a 0.5\% of the accretion
luminosity (efficiency of 10\%) is reprocessed in the disc.

Reduction of $d \ln J /dt$ by factor 2 compared to
Eq.~(\ref{eq:msw}) does not change the results significantly (see
Paper I) and points to the necessity of stronger reduction of AML
for semidetached systems.

\subsection{Reduced angular momentum loss?}
\label{sec:nomsw}

The problem of production of unobserved persistent low-mass
black-hole X-ray binaries under ``standard'' assumptions was noted
before e.g. by \citet[][]{kkb96,ef98,mnl99,iv_kal06}.
\citet{sku72} ``law'' on which \citet{vz81} model is based, is
apparently in conflict with observational data on rotation
velocities in young open clusters
\citep{2002ASPC..261...11C,andron03}. According to the latter study,
the time-scale of rotation braking is two orders of magnitude longer
than the one based on Skumanich law. Necessity for ``weaker'' than
Verbunt \& Zwaan AML by MSW was suggested e.g. by
\citet[][]{ham88,ivtaam03} for cataclysmic variables and by
\citet{vanpar96} for SXT.

In Paper I we suggested, in line with
observational evidence mentioned above, that magnetic braking
operates on a much reduced scale (as compared with \citet{vz81}
prescription) or that it does not operate at all in the
semidetached systems with black-hole accretors. As a test of this
hypothesis, we computed a population of LMBHB under assumption that
MSW is not operating once the RLOF occurs.

Since the mechanism that brings the systems to the RLOF is the same
both in ``standard'' and in ``no-MSW'' cases, the number of
semidetached systems is the same in both models ($\simeq 14000$),
but in the latter case, because of a weaker AML, there are currently
in the Galaxy about 5000 LMBHB that evolve to shorter periods and
have $q \geq 0.02$ (compared to $\simeq 3200$ in the former case).
All these systems are transient according to the criterion
(\ref{eq:mcrit}). This population is plotted in the lower panel of
Fig.~\ref{fig:dim}.

Our estimate of the number of SXT apparently exceeds the estimates
based on observations. Assuming that their outbursts peak on average
at 0.1~$L_{\rm edd}$ in X-rays, all of these transients should be
seen in outburst by the \textit{RXTE ASM}, regardless of their
location in the Galaxy. Matching the observed discovery rate of SXT
would require the outbursts to be sub-Eddington at maximum and/or
recurrence times to be $\apgt 100$~years for most sources.

In the ``no-MSW'' case, donor-stars in the LMBHB have $X_c \apgt
0.35$ at RLOF, while in the ``standard'' case all systems which did
not exhaust hydrogen in the core to the instant of the RLOF evolve
to the shorter$P_{\rm orb}$. This also means that the range of
initial periods of systems in ``no MSW'' case is smaller: it extends
to $\approx35$~hr.

\begin{table}[t!]
\flushleft
\caption[]{Observed short-period low-mass binaries.}
\begin{tabular}{llcccc}
\hline
 &Object & $\per,$ & $Sp$ & $q$ & $M_{\rm bh},$ \\
 & & hr & & & \ms \\
\hline
1& XTE J1118+480 & 4.10 & K5V-M1V &0.044 - 0.035 & 6.5 - 7.2\\
2&GRO J0422+32 & 5.09 & ${\rm M2\pm2V}$ &0.313 - 0.008 &3.7 - 5.0 \\
3&GRS 1009-45 & 6.84 & G5V-M0V & 0.159 - 0.125& 3.6 - 4.7\\
4&XTE J1650-500 & 7.68 & K4 V &  0.1 & 3.4 - 7.9 \\
5&A0620-00 & 7.75 & K3V-K7V & 0.075 - 0.055& 8.7 - 12.9\\
6&GS 2000+25 & 8.28 & K3V-K6V &0.053 - 0.035 & 7.2 - 7.8\\
7&XTE J1859+226 & 9.12 & G5V-K0V & & \\
8&GRS 1124-68 & 10.39 & K3V-K7V &0.208 - 0.114& 6.5 - 8.2 \\
9&H 1705-25 & 12.50 & K3V-K7V & $<0.053$ &5.6 - 8.3 \\
\hline
\end{tabular}
\label{tab:obs}
\end{table}

\subsection{Observational parameters of short-period LMBHB}
\label{sec:obs}

In Table 1 we summarized the main parameters of short orbital period
SXT that are important from evolutionary point of view. The range
of the spectral types assigned to secondaries is based on published
data, see for references Paper II; the values of mass ratios of
components $q=M_2/M_{\rm bh}$ and the estimates of black-hole mases
are taken from \citet{2003IAUS..212..365O,2004ApJ...616..376O}.

\subsubsection{Effective temperatures and masses of donors}
\label{sec:tem}

\begin{figure}
  \includegraphics[angle=-90,width=\columnwidth]{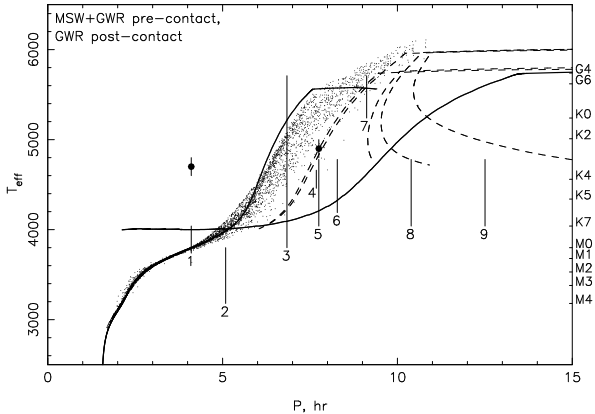}
   \caption[]{Effective temperatures of donors in model population for ``no-MSW'' case (dots). Vertical lines mark the ranges of \te\ of donor-stars in observed SXT corresponding to the ranges of the estimates of their spectral types (Table~\ref{tab:obs}). Systems are annotated according to their numbers in Table~\ref{tab:obs}. Large filled circles give $T_{\rm eff}$ of donors derived from the fits to the synthetic spectra. Dashed lines are evolutionary tracks for $M_0$=1.0 and 1.1\,\ms\ stars with 4\,\ms\ companions . Heavy solid lines to the left and right, respectively, show ``limiting'' tracks for (1+12)\,\ms, $P_0=9.6$ hr system in which MSW does not operate after RLOF and (1+4)\,\ms, $P_0=45.6$ hr system with MSW operating after RLOF (see \S~\ref{sec:tem} for data on tracks and discussion). $Sp - T_{\rm eff}$ relation used in the paper is shown at the right border of the coordinate box.
              }
     \label{fig:pte}
\end{figure}

\begin{figure}
  \includegraphics[angle=-90,width=\columnwidth]{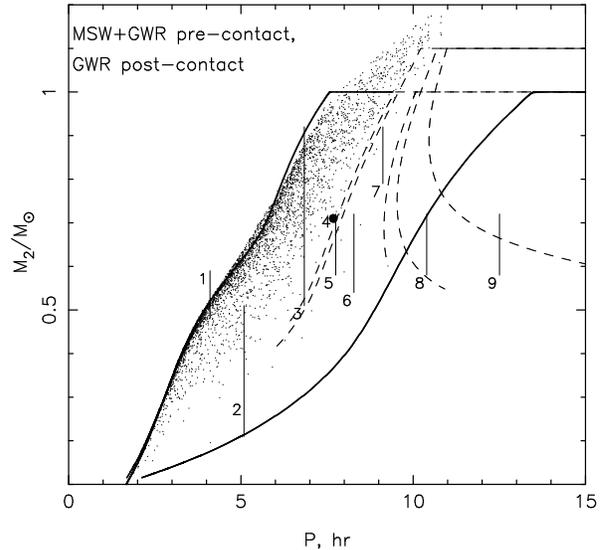}
   \caption[]{Masses of donor-stars in the modeled population for ``no-MSW'' case. Vertical lines show the ranges of $M_2$ corresponding to the range of spectral types of donors in observed SXT (Table \ref{tab:obs}). Dashed and continuous lines are the same evolutionary tracks as in Fig.~\ref{fig:pte}.
For XTE~J1650-500 (at \per=7.88 hr) there is only one determination of the spectrum - K4V, but the same authors mention that the next best fits are G5V and K2III; for this reason we show by solid circle the lower limit for $M_2$ in this system.}
     \label{fig:pm2}
\end{figure}

The essential information available on the companion-stars in SXT is
their spectral types. Determination of the latter is a challenging
task since the emission of the cool star is contaminated by the
radiation from the accretion disc and the hot spot where the
accretion stream hits the disc's edge \citep{2006csxs.book..215C}.
Different methods of \textit{Sp}-determination result in a scatter
of assigned spectra that may amount to several subtypes
(Table~\ref{tab:obs}). In the absence of direct determinations of
such parameters of the stars as their effective temperatures and
masses (for a few exceptions see below), one is forced to use
$Sp-T_{\rm eff}$ and $Sp-M$ relations. We adopted for the present
study $Sp-T_{\rm eff}$ relation by Tokunaga and $Sp-M$ relation by
Schmidt-Kaler given in \citet{2000asqu.book.....C}. In
Figs.~\ref{fig:pte} and \ref{fig:pm2} we plot, respectively,
effective temperatures of the donors and their masses in model
population vs. orbital periods of the systems. In the absence of
objective criteria for discrimination of reliable vs. non-reliable
$Sp$-determinations, we compare model data with the ranges of \te\
and $M_2$ inferred from the spectra.

Within uncertainties of the $Sp$-determinations and conversion $Sp
- \te$ the model satisfactorily reproduces \te\ and $M_2$ of the
donors in the LMBHB with $P_{\rm orb} \lta 9$\ hr. Several systems
-- XTE~1650-500, A0620-00, and GS 2000+25 -- seem to be located
below the ``populated'' areas of the plots. However, we should note
the following. In order to avoid interpolation between pre-computed
tracks that evolve upon RLOF continuously to shorter periods and
tracks that turn to longer periods immediately or change the
direction of evolution in $P_{\rm orb}$, we restricted ourselves by
binaries that evolve to short\per. For a system with given $M_{10}$
and $M_{20}$ the direction of evolution changes quite abruptly, over
a narrow range of initial $\Delta \per \aplt 0.1$\ day. For a given
combination of $M_{10}$ and $M_{20}$ a ``gap'' between tracks
evolving in different directions forms. This is seen from
Fig.~\ref{fig:pte} where we plot evolutionary tracks for (1+4)\,\ms,
$P_0$=1.4 and 1.5 day and for (1.1+4)\,\ms, $P_0$=1.3, 1.4, and 1.45
day systems. But since there is a continuity in the initial
parameters of the systems, the ``gap'' has to be filled by the
systems that start RLOF in the well populated area in the $ P_0 -
M_{20}$ diagram (Fig.~\ref{fig:p0m20}). Thus it is at least
qualitatively clear that the origin of short-period LMBHB may be
explained within paradigm of strong reduction of magnetic braking in
systems with donors overflowing Roche lobes.

In our model we reduced AML by MSW to zero. In reality some AML by MSW can be still
operating.
We plot in Figs.~\ref{fig:pte} and \ref{fig:pm2} two ``limiting'' tracks: for
(1+12)\,\ms, $P_{\rm orb,0}=9.6$ hr in which the donor overflows Roche lobe
almost unevolved
and MSW is absent after RLOF and for (1+4)\,\ms, $P_{\rm orb,0}=45.6$ hr
in which donor has $X_c \simeq 10^{-4}$ at RLOF and MSW continues
to operate. Crudely,
model populations with MSW and without MSW that evolve to short \per\
have to be located between
these two limiting curves. There is a contribution of lower and
higher mass systems of course as we plotted tracks for 1\ms\ donors for simplicity.
Adding some AML to our model will shift the population to
the right, giving a better agreement with observations while still
not producing stable luminous X-ray sources. From Fig. 6 in Paper I it is
seen that such an adding of AML will influence mainly the long-period
systems.

Our computations for the ``no-MSW'' model show
that LMBHB evolve to longer periods if $X_c \lta 0.35$ at RLOF. For
instance a 1.1\ms\ donor with $\per_0=1.4$\ (middle curve)
spends in the RLOF-state almost 10~Gyr, out of which about 5~Gyr it
evolves to longer periods. This provides a possibility of
explaining SXTs with periods $>$ 9-10 hr.

\subsubsection{Mass-transfer rates}
\label{sec:dmdt}

Our model predicts the distribution of SXT over mass-transfer rates
\md\ that can also be compared to \md\ deduced from observations of
SXT sources.

Usually, \md\ for SXT are evaluated
by
 dividing the mass accreted during outburst by the recurrence time
\citep[see e.g.][]{wp96}. Among short-period SXT the
recurrence time is known only for A0620-00 (about 60 years) and
4U~1543-47 (about 10 years). For the other systems one can only obtain
upper limits of \md\ since only one outburst has been seen
in the X-ray observations epoch (30 years are usually assumed).
 These estimates of \md,
derived from the data on outburst parameters from \citet{csl97} and
for KV UMa and V406 Vul based of observations presented in
\citet{chaty1118} and \citet{hynes1859} respectively, are
shown as plusses in
Figs.~\ref{fig:dim} and \ref{fig:mnl}.

For the ``standard'' model all above-described \md-estimates are
inconsistent with the model (two upper panels of
Fig.~\ref{fig:dim}). Even if only GWR acts, $\md \sim
10^{-12}$\,\myr\ cannot be explained if one assumes that they
represent {\sl secular} values (lower panel of Fig.~\ref{fig:dim}).
It would take post-minimum period systems several Hubble times to
reach $\per \apgt 4$ hr and have a secular $\md \sim
10^{-12}$\,\myr. If the extremely low \md\ deduced for short period
SXTs were due to downward fluctuations from an intrinsically high
secular
 \md,
one would still have to explain why the bright counterparts are not
observed (\S~\ref{sec:stable}). Furthermore, some systems
would then necessarily have higher rates than their secular value,
hence there should be even more persistent systems.

In Paper I we have shown that in the ``standard'' model systems with
$P_{\rm orb} \approx 8 - 10$\, hr may be consistent with systems
where donors overflow Roche lobes extremely close to TAMS, but then
one needs to assume some very special initial distribution of
binaries over orbital separations which will after several
evolutionary stages with mass transfer and mass and momentum loss
from the system and supernova explosion, result in concentration of
zero-age black-hole and main-sequence-star systems just at the
desired very narrow range of separations. On the other hand, if a
small amount of AML is added to the alternative ``no-MSW'' model,
the latter will become consistent with recurrence time based
estimates of \md\ for observed $P_{\rm orb} \approx 8 - 10$\, hr
systems, as shows comparison of two lower panels of
Fig.~\ref{fig:dim}.

The longest period SXT of the ``short-period'' group, 4U1543-47 
($\per=27$\,hr; it is  not shown in the Figures) is, in
principle, consistent with the evolution of ``standard'' model systems with donors
overflowing Roche-lobes after formation of He-cores. But again,
these binaries evolve through the Hertzsprung gap extremely fast, and
 the probability of observing such systems is very low
  \citep[see ][]{kolb_soft98}. In the ``no-MSW'' model, formation of the
systems similar to 4U1543-47 is much easier explained, since to the
range of periods about 13 hr slowly enough evolve binaries with
$M_{20} \simeq 1$\,\ms\ donors which overfilled Roche lobes having
$X_c \aplt 0.35$.

\subsubsection{Truncated discs \label{sec:mnlpar}}

It is, however, plausible that the \md\ in short-period SXTs are
much higher than \md\ derived above. These \md\ were obtained
treating accretion disc as a reservoir which during quiescence is
filled up to a critical value at which the outburst is triggered and
the disc emptied. The implicit assumption is that the reservoir is
not leaky, i.e. that the accretion rate at the disc's inner edge is
much smaller than \md. However, the inner truncation of the discs is
required to explain observed quiescent luminosities \citep[e.g.
][]{l96,letal96}. The same may be true for non-quiescent discs too
\citep[see e.g. ][]{done_gier06}. \citet{dubhamlas01} showed that
the disc instability model can reproduce the observed X-ray
light-curves only if discs in SXTs are truncated and irradiated. The
truncated (or ``leaky'') disc paradigm is now commonly used in
describing the observed timing and spectral properties of X-ray
LMBHB \citep[][ and references therein]{mclrbh}.

For systems with {\sl non-stationary} quiescent accretion discs \md\
can be estimated as \citep{lasota01}
\begin{equation}
{\dot M \approx \frac{\epsilon M_{\rm D, max}}{t_{\rm
recc}}+ \dot M_{\rm in}}
 \label{eq:mdot}
\end{equation}
where
\begin{equation}
M_{\rm D, max}=2.7 \times 10^{21} \alpha^{-0.83} \left( \frac{M_1}
{\rm M_\odot} \right)^{-0.38} R_{d,10}^{3.14}, \label{eq:discmass}
\end{equation}
is the maximum quiescent-disc mass (in ${\rm g}$) and $\epsilon= \Delta M_{\rm
D}/M_{\rm D, max}$ -- the fraction of the disc's mass lost during
outburst; $\alpha$ is the kinematic viscosity parameter, $R_{d,10}$ is
the disc's outer radius in $10^{10}$ cm and $\dot M_{\rm in}$ is the accretion rate
at the disc's inner edge. The usual, non-leaky disc estimates,
neglect $\dot M_{\rm in}$.
However, in SXTs $\dot M_{\rm in}$ is the dominant term as
argued by \citet{mnl99}.

\begin{figure}[ht!]
\includegraphics[angle=-90,width=0.48\textwidth]{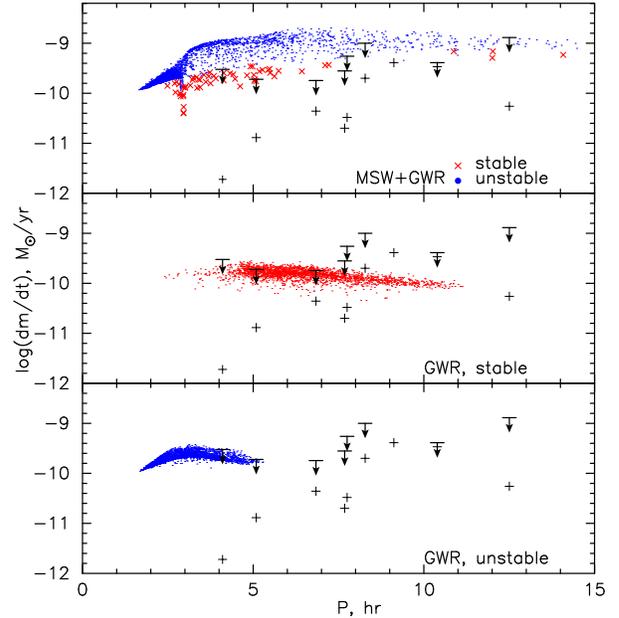}
\caption[]{Mass transfer rates vs. orbital period for the model
population of semidetached low-mass binaries with assumed black hole
accretors. Upper panel -- model for Verbunt-Zwaan assumptions on
AML via MSW. Crosses mark systems that are stable according to
Menou et al. criterion for instability of cold discs, while dots
mark unstable systems. Middle panel shows position of systems with discs
stable according to Menou et al. criterion in the model in which MSW
does not operate in semidetached systems. Lower panel -- systems with unstable discs in model without MSW.
Arrows and plusses have the same meaning as in Fig.~\ref{fig:dim}.}
\label{fig:mnl}
\end{figure}

According to the DIM, the disc is in cold thermal equilibrium if
accretion rate at all annuli satisfies the inequality
\citep{2008arXiv0802.3848L}
\begin{multline}
\md(r)  < \md^-_{\rm crit} \\
\approx2.64\times10^{15}\alpha_{0.1}^{0.01}R_{10}^{
2.58}\left(\frac{M_1}{\ms}\right)^{-0.85}~{\rm g~~s^{-1}},
\label{eq:mcritcold}
\end{multline}
where $\alpha_{0.1}$ -- the viscosity parameter in 0.1. 
Therefore, the mass accretion
rate at the truncation radius ($\dot M_{\rm in}$) must be smaller
than $\md^-_{\rm crit}(r_{\rm in})$. Parameterizing the truncation radius as a fraction
of the circularization radius $r_{\rm in}=f_t r_{\rm
circ}$ \citep[where $f_t < 0.48$, see][]{mnl99} one gets (in ${\rm \ms\,yr^{-1}}$)
\begin{multline}
\md_{\rm max}\aplt2.5\times 10^{-7} \\
\left[\left(1+q\right)^{1/3}\left(0.5-0.227 \log q\right)\right]^{10.32} P_d^{1.72}f_t^{2.58}, 
\label{eq:dmmax}
\end{multline}
where period is in days. Inserting in Eq. (\ref{eq:dmmax}) the lower
limits on $q$ from Table \ref{tab:obs} and assuming $f_t = 0.48$,
one obtains the upper limits to the mass-transfer rates shown by
arrows in Figs.~\ref{fig:dim} and \ref{fig:mnl}. One can therefore
conclude that a substantially reduced MSW AML would be consistent
with both the stability properties and \md\
 of SXTs.

\section{Cold, stable systems}
\label{sec:cold}

Possible problem of excess of SXT in our model mentioned in \S~\ref{sec:stable} may be resolved in the model
suggested by \citet{mnl99} (henceforth, MNL).
MNL pointed
out that if the truncation radii were slightly larger than estimated
to fit observations of quiescent SXTs the discs would be globally
stable, since their accretion rates are lower
than the critical value given by Eq. (\ref{eq:mcritcold})
corresponding to cold stable equilibria.

Figure \ref{fig:mnl} shows the distribution of stars in two
population models with respect to stability criterion suggested by
MNL under assumption the truncation occurs at $r_{in}= 0.48 r_{\rm
circ}$. In the ``standard'' case with MSW AML all ($\simeq 2900$),
but very few ($\simeq 60$), systems are unstable
(Fig.~\ref{fig:mnl} upper panel) vs. MNL-criterion. On the other
hand, in the model without MSW after RLOF there are almost equal
numbers ($\simeq 2500$) of stable and unstable vs.
 MNL-criterion systems
(Fig.~\ref{fig:mnl}, two lower panels). This is because \md\ in the
GWR-only scenario are lower than in the MSW+GWR scenario, thereby
enabling the criterion for a cold disc to be met more easily.
Truncating the disc at R$_{\rm circ}$ would make all transients
``MNL-stable'' for both models. Therefore, it is possible that most
or even all LMBHB are secularly cold/stable systems, erupting from
time to time due to randomly acting factors like variability of
truncation radii or mass transfer rate. The predicted population
size of these systems is too small to contribute significantly to
the large number of faint ($L_X\sim 10^{31}$~erg~s$^{-1}$) hard
X-ray sources seen in the deep \textit{Chandra} exposures toward the
Galactic Center \citep{muno03}.

\section{Conclusion}
\label{sec:concl}

1. We have modeled populations of short-period semidetached close
binary systems containing black holes and low-mass stars under
various assumptions about the AML mechanism.

We have found that the "standard" model (based on \citet{sku72}
stellar rotation braking law prescription for AML via magnetic
braking) produces unobserved persistent LMBHB, while a model with
pure GWR AML produces only transient systems. In the latter model it
is possible to reproduce, within uncertainty of observations, the
number of LMBHB in the Galaxy, the effective temperatures and masses
of the donors in these systems (as inferred from the spectra of the
latter). Mass-transfer rates in this model are consistent with the
upper limits on \md, if quiescent accretion discs are truncated and,
hence, leaky. All this suggests that the strength of MSW AML
should be strongly reduced compared to the ``standard'' model.

In the model without strong MSW it is unnecessary to invoke systems
with the donors that are almost at TAMS or already left the latter
($X_c \lesssim 0.01$) in order to explain observed orbital periods
of SXT and their mass-exchange rates. More, population of
short-period ($\per \aplt 8-9$ hr) is formed by systems that
overflow Roche lobes unevolved or slightly evolved ($X_c > 0.35$),
while evolution of systems with more evolved donors with masses
$\simeq 1\,\ms$ opens possibility to explain the origin of LMBHB
with longer \per.

2. An alternative model suggests that if accretion discs are
truncated in quiescence they could be cold and stable. Then
truncation of discs at radii close to $R_{\rm circ}$ can make discs
in virtually all LMBHB cold and stable. In this case SXT outbursts
perhaps are not related to DIM and are random events that result
from "external factors", like episodes of enhanced mass transfer or
variability in truncation radius.

3. Population synthesis models presented above are obtained under
assumption of a high value of the product of binding energy
parameter of stellar envelopes and expulsion efficiency of common
envelopes -- \al=2. Model with \al=0.5 gives similar results (but
a reduced total number of systems). Further decrease of \al\
results in models that are not consistent with observed SXT. The
issue of \al\ is still controversial \citep[see, e.g.
][]{Taudew,Podsiadlowski&al03} in the sense that there are no strict
criteria for defining binding energy of stellar envelopes and there
is no clear understanding whether sources other than gravitational
energy may contribute to unbinding common envelopes. Apart of common
envelopes, crucial role in determining the possibility of formation
of LMBHB is played by stellar wind mass-loss in all stages of
stellar evolution, since it defines parameters of the system at the
beginning of common envelope stage and immediately before supernova
explosion that gives birth to black hole. If general notions on
mass-loss will not be revised, our results suggest that existence of
short-period low-mass X-ray binaries demands high values of \al\
parameter.

\section{Acknowledgements}
\label{sec:ack} We acknowledge P.P. Eggleton for providing a copy of
his evolutionary code, G. Nelemans and G. Dubus for fruitful cooperation and L. Titarchuk and N. Shaposhnikov for useful remarks.
 LRY acknowledges warm hospitality and support
from Institut d'Astrophysique de Paris, Universit\'e Pierre et Marie
Curie where most of this study was carried out and support 
by Polish
Ministry of Education grant N203 009 31/1466 and Swedish VR grant Dnr
621-2006-3288  which enabled his participation in the
conference.  LRY is
supported by RFBR grant 07-02-00454 and Russian Academy of Sciences
Basic Research Program ``Origin and Evolution of Stars and
Galaxies''.









\begin{thebibliography}{49}
\expandafter\ifx\csname natexlab\endcsname\relax\def\natexlab#1{#1}\fi
\expandafter\ifx\csname url\endcsname\relax
  \def\url#1{\texttt{#1}}\fi
\expandafter\ifx\csname urlprefix\endcsname\relax\def\urlprefix{URL }\fi

\bibitem[{{Andronov} et~al.(2003){Andronov}, {Pinsonneault}, and
  {Sills}}]{andron03}
{Andronov}, N., {Pinsonneault}, M., {Sills}, A., Jan. 2003. {Cataclysmic
  Variables: An Empirical Angular Momentum Loss Prescription from Open Cluster
  Data}. \apj 582, 358--368.

\bibitem[{{Boissier} and {Prantzos}(1999)}]{bp99}
{Boissier}, S., {Prantzos}, N., Aug. 1999. {Chemo-spectrophotometric evolution
  of spiral galaxies - I. The model and the Milky Way}. \mnras 307, 857--876.

\bibitem[{{Charles} and {Coe}(2006)}]{2006csxs.book..215C}
{Charles}, P.~A., {Coe}, M.~J., Apr. 2006. {Optical, ultraviolet and infrared
  observations of X-ray binaries}. In: Lewin, W., van~der Klis, M. (Eds.),
  Compact stellar X-ray sources. Cambridge, pp. 215--265.

\bibitem[{{Chaty} et~al.(2003){Chaty}, {Haswell}, {Malzac}, {Hynes}, {Shrader},
  and {Cui}}]{chaty1118}
{Chaty}, S., {Haswell}, C.~A., {Malzac}, J., {Hynes}, R.~I., {Shrader}, C.~R.,
  {Cui}, W., Dec. 2003. {Multiwavelength observations revealing the evolution
  of the outburst of the black hole XTE J1118+480}. \mnras 346, 689--703.

\bibitem[{{Chen} et~al.(1997){Chen}, {Shrader}, and {Livio}}]{csl97}
{Chen}, W., {Shrader}, C.~R., {Livio}, M., Dec. 1997. {The Properties of X-Ray
  and Optical Light Curves of X-Ray Novae}. \apj 491, 312.

\bibitem[{{Collier Cameron}(2002)}]{2002ASPC..261...11C}
{Collier Cameron}, A., Jan. 2002. {Magnetic activity in low-mass stars: Do the
  brakes come off?} In: {G{\"a}nsicke}, B.~T., {Beuermann}, K., {Reinsch}, K.
  (Eds.), The Physics of Cataclysmic Variables and Related Objects. Vol. 261 of
  Astronomical Society of the Pacific Conference Series. p.~11.

\bibitem[{{Cox}(2000)}]{2000asqu.book.....C}
{Cox}, A.~N., 2000. {Allen's astrophysical quantities, 4th ed.} New York: AIP
  Press; Springer, 2000.~Ed. Arthur N.~Cox.

\bibitem[{de~Kool et~al.(1987)de~Kool, van~den Heuvel, and Pylyser}]{khp87}
de~Kool, M., van~den Heuvel, E. P.~J., Pylyser, E., 1987. An evolutionary
  scenario for the black hole binary a0620-00. \aap 183, 47--52.

\bibitem[{{Done} and {Gierli{\'n}ski}(2006)}]{done_gier06}
{Done}, C., {Gierli{\'n}ski}, M., Jan. 2006. {Truncated disc versus extremely
  broad iron line in XTE J1650-500}. \mnras, 121.

\bibitem[{{Dubus} et~al.(2001){Dubus}, {Hameury}, and {Lasota}}]{dubhamlas01}
{Dubus}, G., {Hameury}, J.-M., {Lasota}, J.-P., Jul. 2001. {The disc
  instability model for X-ray transients: Evidence for truncation and
  irradiation}. \aap 373, 251--271.

\bibitem[{{Dubus} et~al.(1999){Dubus}, {Lasota}, {Hameury}, and
  {Charles}}]{dubhamlas99}
{Dubus}, G., {Lasota}, J., {Hameury}, J., {Charles}, P., Feb. 1999. {X-ray
  irradiation in low-mass binary systems}. \mnras 303, 139--147.

\bibitem[{{Eggleton}(1971)}]{egg71}
{Eggleton}, P.~P., 1971. {The evolution of low mass stars}. \mnras 151, 351.

\bibitem[{Ergma and Fedorova(1998)}]{ef98}
Ergma, E., Fedorova, A., 1998. Evolution of black hole low-mass binaries. \aa
  338, 69--74.

\bibitem[{{Esin} et~al.(2000){Esin}, {Lasota}, and {Hynes}}]{elh}
{Esin}, A.~A., {Lasota}, J.-P., {Hynes}, R.~I., Feb. 2000. {The 1996 outburst
  of GRO J1655-40: disc irradiation and enhanced mass transfer}. \aap 354,
  987--994.

\bibitem[{{Fryer} and {Kalogera}(2001)}]{fryer_kal_bh01}
{Fryer}, C.~L., {Kalogera}, V., Jun. 2001. {Theoretical Black Hole Mass
  Distributions}. \apj 554, 548--560.

\bibitem[{{Hameury} et~al.(1988){Hameury}, {Lasota}, {King}, and
  {Ritter}}]{ham88}
{Hameury}, J.~M., {Lasota}, J.~P., {King}, A.~R., {Ritter}, H., Mar. 1988.
  {Magnetic braking and the evolution of cataclysmic binaries}. \mnras 231,
  535--547.

\bibitem[{{Hobbs} et~al.(2005){Hobbs}, {Lorimer}, {Lyne}, and
  {Kramer}}]{hobbs2005}
{Hobbs}, G., {Lorimer}, D.~R., {Lyne}, A.~G., {Kramer}, M., Jul. 2005. {A
  statistical study of 233 pulsar proper motions}. \mnras 360, 974--992.

\bibitem[{{Hynes} et~al.(2002){Hynes}, {Haswell}, {Chaty}, {Shrader}, and
  {Cui}}]{hynes1859}
{Hynes}, R.~I., {Haswell}, C.~A., {Chaty}, S., {Shrader}, C.~R., {Cui}, W.,
  Mar. 2002. {The evolving accretion disc in the black hole X-ray transient XTE
  J1859+226}. \mnras 331, 169--179.

\bibitem[{{Ivanova} and {Kalogera}(2006)}]{iv_kal06}
{Ivanova}, N., {Kalogera}, V., Jan. 2006. {The Brightest Point X-Ray Sources in
  Elliptical Galaxies and the Mass Spectrum of Accreting Black Holes}. \apj
  636, 985--994, arXiv:astro-ph/0506471.

\bibitem[{{Ivanova} and {Taam}(2003)}]{ivtaam03}
{Ivanova}, N., {Taam}, R.~E., Dec. 2003. {Magnetic Braking Revisited}. \apj
  599, 516--521.

\bibitem[{{King} et~al.(1996){King}, {Kolb}, and {Burderi}}]{kkb96}
{King}, A.~R., {Kolb}, U., {Burderi}, L., Jun. 1996. Black hole binaries and
  x-ray transients. \apjl 464, L127.

\bibitem[{{Kolb}(1998)}]{kolb_soft98}
{Kolb}, U., Jun. 1998. Soft x-ray transients in the hertzsprung gap. \mnras
  297, 419--426.

\bibitem[{{Kroupa} et~al.(1993){Kroupa}, {Tout}, and {Gilmore}}]{ktg03}
{Kroupa}, P., {Tout}, C.~A., {Gilmore}, G., Jun. 1993. {The distribution of
  low-mass stars in the Galactic disc}. \mnras 262, 545--587.

\bibitem[{Landau and Lifshitz(1971)}]{ll71}
Landau, L.~D., Lifshitz, E.~M., 1971. "Classical theory of fields", 3rd
  Edition. Pergamon, Oxford.

\bibitem[{{Lasota}(1996)}]{l96}
{Lasota}, J.~P., 1996. {Mechanisms for Dwarf Nova Outbursts and Soft X-ray
  Transients (A Critical Review)}. In: IAU Symp. 165: Compact Stars in
  Binaries. p.~43.

\bibitem[{{Lasota}(2001)}]{lasota01}
{Lasota}, J.-P., Jun. 2001. {The disc instability model of dwarf novae and
  low-mass X-ray binary transients}. New Astronomy Review 45, 449--508.

\bibitem[{{Lasota} et~al.(2008){Lasota}, {Dubus}, and
  {Kruk}}]{2008arXiv0802.3848L}
{Lasota}, J.-P., {Dubus}, G., {Kruk}, K., Feb. 2008. {Stability of helium
  accretion discs in ultracompact binaries}. ArXiv e-prints 0802.3848.

\bibitem[{{Lasota} et~al.(1996){Lasota}, {Narayan}, and {Yi}}]{letal96}
{Lasota}, J.-P., {Narayan}, R., {Yi}, I., Oct. 1996. {Mechanisms for the
  outbursts of soft X-ray transients.} \aap 314, 813--820.

\bibitem[{{McClintock} and {Remillard}(2006)}]{mclrbh}
{McClintock}, J.~E., {Remillard}, R.~A., Apr. 2006. {Black hole binaries}. In:
  Lewin, W., van~der Klis, M. (Eds.), Compact stellar X-ray sources. Cambridge,
  pp. 157--213.

\bibitem[{{Menou} et~al.(1999){Menou}, {Narayan}, and {Lasota}}]{mnl99}
{Menou}, K., {Narayan}, R., {Lasota}, J., Mar. 1999. A population of faint
  nontransient low-mass black hole binaries. \apj 513, 811--826.

\bibitem[{{Muno} et~al.(2003){Muno}, {Baganoff}, {Bautz}, {Brandt}, {Broos},
  {Feigelson}, {Garmire}, {Morris}, {Ricker}, and {Townsley}}]{muno03}
{Muno}, M.~P., {Baganoff}, F.~K., {Bautz}, M.~W., {Brandt}, W.~N., {Broos},
  P.~S., {Feigelson}, E.~D., {Garmire}, G.~P., {Morris}, M.~R., {Ricker},
  G.~R., {Townsley}, L.~K., May 2003. {A Deep Chandra Catalog of X-Ray Point
  Sources toward the Galactic Center}. \apj 589, 225--241.

\bibitem[{{Nelemans} et~al.(2004){Nelemans}, {Yungelson}, and {Portegies
  Zwart}}]{nyp04}
{Nelemans}, G., {Yungelson}, L.~R., {Portegies Zwart}, S.~F., Mar. 2004.
  {Short-period AM CVn systems as optical, X-ray and gravitational-wave
  sources}. \mnras 349, 181--192.

\bibitem[{Nelemans et~al.(2001)Nelemans, Yungelson, Portegies~Zwart, and
  Verbunt}]{nyp+01}
Nelemans, G., Yungelson, L.~R., Portegies~Zwart, S.~F., Verbunt, F., 2001.
  Population synthesis for double white dwarfs. i detached systems. \aap 365,
  491 -- 507.

\bibitem[{{Orosz}(2003)}]{2003IAUS..212..365O}
{Orosz}, J.~A., 2003. {Inventory of black hole binaries}. In: {van der Hucht},
  K., {Herrero}, A., {Esteban}, C. (Eds.), A Massive Star Odyssey: From Main
  Sequence to Supernova. Vol. 212 of IAU Symposium. p. 365.

\bibitem[{{Orosz} et~al.(2004){Orosz}, {McClintock}, {Remillard}, and
  {Corbel}}]{2004ApJ...616..376O}
{Orosz}, J.~A., {McClintock}, J.~E., {Remillard}, R.~A., {Corbel}, S., Nov.
  2004. {Orbital Parameters for the Black Hole Binary XTE J1650-500}. \apj 616,
  376--382.

\bibitem[{{Paczy\'nski}(1990)}]{pac90}
{Paczy\'nski}, B., Jan. 1990. {A test of the galactic origin of gamma-ray
  bursts}. \apj 348, 485--494.

\bibitem[{{Podsiadlowski} et~al.(2003){Podsiadlowski}, {Rappaport}, and
  {Han}}]{Podsiadlowski&al03}
{Podsiadlowski}, P., {Rappaport}, S., {Han}, Z., May 2003. {On the formation
  and evolution of black hole binaries}. \mnras 341, 385--404.

\bibitem[{Portegies~Zwart and Verbunt(1996)}]{pv96}
Portegies~Zwart, S.~F., Verbunt, F., 1996. Population synthesis of high-mass
  binaries. \aap 309, 179--196.

\bibitem[{Portegies~Zwart and Yungelson(1998)}]{py98}
Portegies~Zwart, S.~F., Yungelson, L.~R., 1998. Formation and evolution of
  binary neutron stars. \aap 332, 173.

\bibitem[{{Remillard} and {McClintock}(2006)}]{2006ARA&A..44...49R}
{Remillard}, R.~A., {McClintock}, J.~E., Sep. 2006. {X-Ray Properties of
  Black-Hole Binaries}. \araa 44, 49--92, arXiv:astro-ph/0606352.

\bibitem[{{Romani}(1998)}]{romani1998}
{Romani}, R.~W., May 1998. {A census of low mass black hole binaries}. \aap
  333, 583--590.

\bibitem[{{Shakura} and {Sunyaev}(1973)}]{shaksun73}
{Shakura}, N.~I., {Sunyaev}, R.~A., 1973. {Black holes in binary systems.
  Observational appearance.} \aap 24, 337--355.

\bibitem[{{Skumanich}(1972)}]{sku72}
{Skumanich}, A., Feb. 1972. {Time Scales for CA II Emission Decay, Rotational
  Braking, and Lithium Depletion}. \apj 171, 565.

\bibitem[{Tauris and Dewi(2001)}]{Taudew}
Tauris, T., Dewi, J. D.~M., Apr. 2001. On the binding energy parameter of
  common envelope evolution. \aap 369, 170--173.

\bibitem[{{van Paradijs}(1996)}]{vanpar96}
{van Paradijs}, J., Jun. 1996. {On the Accretion Instability in Soft X-Ray
  Transients}. \apjl 464, L139.

\bibitem[{{Verbunt} and {Zwaan}(1981)}]{vz81}
{Verbunt}, F., {Zwaan}, C., Jul. 1981. Magnetic braking in low-mass x-ray
  binaries. \aap 100, L7--L9.

\bibitem[{Webbink(1984)}]{web84}
Webbink, R.~F., 1984. Double white dwarfs as progenitors of r coronae borealis
  stars and type i supernovae. \apj 277, 355.

\bibitem[{White and van Paradijs(1996)}]{wp96}
White, N.~E., van Paradijs, J., 1996. The galactic distribution of black hole
  candidates in low-mass x-ray binary systems. \apj 473, L25--L29.

\bibitem[{{Yungelson} et~al.(2006){Yungelson}, {Lasota}, {Nelemans}, {Dubus},
  {van den Heuvel}, {Dewi}, and {Portegies Zwart}}]{yungelson_bh06}
{Yungelson}, L.~R., {Lasota}, J.-P., {Nelemans}, G., {Dubus}, G., {van den
  Heuvel}, E.~P.~J., {Dewi}, J., {Portegies Zwart}, S., Aug. 2006. {The origin
  and fate of short-period low-mass black-hole binaries}. \aap 454, 559--569.

\end{thebibliography}

\end{document}